\documentclass[letterpaper]{article} 
\usepackage{aaai24}  
\usepackage{times}  
\usepackage{helvet}  
\usepackage{courier}  
\usepackage[hyphens]{url}  
\usepackage{graphicx} 
\usepackage{natbib}  
\usepackage{caption} 
\usepackage{algorithm}
\usepackage{algorithmic}

\usepackage{amsmath}
\usepackage{subcaption}
\usepackage{enumitem}

\usepackage{newfloat}
\usepackage{listings}
\DeclareCaptionStyle{ruled}{labelfont=normalfont,labelsep=colon,strut=off} 
\floatstyle{ruled}
\newfloat{listing}{tb}{lst}{}
\floatname{listing}{Listing}
\title{Benchmarking Private Population Data Release Mechanisms: \\ Synthetic Data vs. TopDown}
\author {
    Aadyaa Maddi,
    Swadhin Routray,
    Alexander Goldberg,
    Giulia Fanti
}
\affiliations {
    Carnegie Mellon University\\
    amaddi@andrew.cmu.edu, sroutray@andrew.cmu.edu, akgoldbe@cs.cmu.edu, gfanti@andrew.cmu.edu
}

\begin{document}

\maketitle

\begin{abstract}
Differential privacy (DP) is increasingly used to protect the release of hierarchical, tabular population data, such as census data \cite{abowd20222020}.
A common approach for implementing DP in this setting is to release noisy responses to a predefined set of queries. For example, this is the approach of the TopDown algorithm \cite{fioretto2021differential} used by the US Census Bureau.
Such methods have an important shortcoming: 
they cannot answer queries for which they were not optimized.
An appealing alternative is to generate DP synthetic data, which  is drawn from some generating distribution. 
Like the TopDown method, synthetic data can also be optimized to answer specific queries \cite{liu2022private}, while also allowing the data user to later submit arbitrary queries over the synthetic population data.
To our knowledge, there has not been a head-to-head empirical comparison of these approaches. 
This study conducts such a comparison between the TopDown algorithm and private synthetic data generation to determine how accuracy is affected by query complexity, in-distribution vs. out-of-distribution queries, and privacy guarantees. Our results show that for in-distribution queries, the TopDown algorithm achieves significantly better privacy-fidelity tradeoffs than any of the synthetic data methods we evaluated; for instance, in our experiments, TopDown achieved at least $20\times$ lower error on counting queries than the leading synthetic data method at the same privacy budget \cite{mckenna2021winning}. 
Our findings suggest guidelines for practitioners and the synthetic data research community. 
\end{abstract}

\section{Introduction}

Countries routinely compile and publish census data on residents for government operations. For example, personally identifiable information like sex, age, race, and ethnicity, are some of the features collected in the US Census. 
Due to the sensitive nature of this data, countries release tables of aggregate statistics instead of publishing raw data in the clear. These statistics must be accurate because of critical downstream tasks like ``apportionment, redistricting, allocation of funds, public policy, and research" \cite{abowd20222020}. For example, population counts from the United States Census determine the number of elected officials each state can have in the House of Representatives \cite{cohen2022census}. 

However, releasing aggregate statistics is not enough to prevent attackers from reconstructing individual rows of a dataset with near-perfect accuracy
\cite{cohen2020linear,narayanan2008robust}. 
To protect against this privacy risk, population statistics have been released with formal differential privacy (DP) guarantees, most famously by the US Census \cite{abowd20222020,dwork2006calibrating}. 
Two popular approaches to releasing private data with differential privacy are: (1) adding calibrated noise to aggregate statistics before releasing them publicly \cite{abowd20222020}, and (2) privately approximating the underlying data distribution from the raw data and generating synthetic data from this distribution to be released publicly \cite{liu2022private,sweden2023pop}. 

Directly applying many existing DP mechanisms to census data is not possible because of constraints arising from its hierarchical nature and applicable administrative requirements. For example, the US Census divides the country into administrative units called census blocks. These blocks are combined to release counts at a six-level hierarchy: nation, state, county, tract, block group, and block \cite{abowd20222020}. Private statistics for the census need to be non-negative integers and consistent at each geographical level. For example, the sum of the populations of all cities in a state must equal the population of the state. For applications like apportionment, counts must be invariant or match the ground truth values at some geographical levels \cite{abowd20222020}. 
This has given rise to two solutions:

\paragraph{Approach 1: Private statistic release.}
The US Census addresses the challenges of hierarchical data with the TopDown algorithm, which releases aggregate statistics while respecting hierarchical constraints \cite{abowd20222020}. The TopDown algorithm samples noise from a geometric distribution and adds it to each aggregate query result. Then, a post-processing phase guarantees that the published counts are non-negative integers and that the total counts at each geographical level are equal. \citet{fioretto2021differential} propose a more efficient version of this algorithm, which we use in our evaluation. 

\paragraph{Approach 2: Private synthetic data.}
DP synthetic data generation instead  generates a synthetic population on which queries can be computed.  
This approach trivially meets hierarchical constraints because queries are computed from (synthetic) individuals.
However, it is challenging to accurately estimate the distribution of census data due to the high dimensionality of the dataset. There are various methods for addressing this challenge. For instance, the authors in \cite{liu2022private} propose modeling the underlying distribution as a mixture of hierarchical product distributions (HPD) across features. Their approach is motivated by the fact that records can be nested under group-level features (such as multiple individuals comprising one census block). In this work, we evaluate both fixed distribution modeling (HPD-Fixed) and generative modeling (HPD-Gen) approaches introduced in \cite{liu2022private}. 

An alternative Maximum Spanning Tree (MST) approach  \cite{mckenna2021winning} builds on the Chow-Liu algorithm \cite{chow1968approximating} to privately infer an underlying probabilistic graphical model, and generates synthetic data from this distribution.
We compare against both algorithms because to our knowledge, HPD (-Fixed and -Gen) are the only private synthetic data generation algorithms optimized for hierarchical data; on the other hand, MST is a general-purpose synthetic data algorithm for tabular data, but was shown to outperform other synthetic data algorithms in a recent benchmarking study \cite{tao2021benchmarking}.

Private synthetic data offers flexibility in selecting which queries must be answered. The TopDown algorithm, on the other hand, requires prior knowledge of the queries. We distinguish between these two types of queries: in-distribution and out-of-distribution. \emph{In-distribution} queries refer to those on which the synthetic dataset and TopDown algorithm were trained, whereas \emph{out-of-distribution} queries are unknown beforehand. 

\paragraph{Contributions.} 
Our contributions are threefold. 
\begin{enumerate}[nosep]
    \item We provide the first empirical comparison (to our knowledge) of state-of-the-art methods for tabular, hierarchical private data release. Our experiments show that the TopDown algorithm performs significantly better than PrivSyn (HPD-Fixed), PrivSyn (HPD-Gen), and MST in answering in-distribution queries, regardless of query complexity and desired privacy guarantees (though it cannot answer out-of-distribution queries).
    \item We evaluate the quality of synthetic datasets using the metrics introduced in \cite{tao2021benchmarking}. We use this evaluation to study which properties of the underlying data distribution the synthetic data algorithms learn best. We find that MST is more effective than HPD-Gen and HPD-Fixed in preserving the ground truth distributions of individual features, pairs of features, and correlations between features, even though HPD algorithms are trained on in-distribution queries.
    \item We present guidelines for dataset providers to select a data release algorithm. Overall, we recommend that providers use the TopDown algorithm for in-distribution queries, and opt for synthetic data from the MST algorithm when the queries are unknown beforehand.
\end{enumerate}

\section{Private Tabular Data Release}

\subsection{Problem Formulation} 
A dataset provider (e.g., country) collects information on $n$ individuals and wants to release aggregate statistics to approved third parties. Denote the dataset as $\mathbf{D} = \{(p_i, x_{i,1}, \dots, x_{i,m}, \dots, x_{i,M}, r_i)\}_{i=1}^{n}$ containing $n$ records, where each record is a tuple $\in \mathbf{P} \times \mathbf{X}^{M} \times \mathbf{R}$. Here, $p_i \in \mathbf{P}$ is a unique identifier for person $i$ in the dataset. Each person has an $M$-dimensional feature vector $(x_{i,1}, \dots, x_{i,m},\dots, x_{i,M}) \in \mathbf{X}^{M}$, where the $m$\textsuperscript{th} feature value for person $i$, i.e., $x_{i, m}$ (e.g., is employed) is sampled from the corresponding domain $\mathbf{X}_m$ (e.g. employment status). The geographical region that person $i$ resides in is given by $r_i \in \mathbf{R}$ (e.g., county name).

Following the notation introduced in \citep{fioretto2021differential}, the dataset provider has a region hierarchy defined using a tree $\mathcal{T}$ with $L$ levels. The dataset provider uses this tree to release aggregate statistics grouped by geographical region. Each level of this tree corresponds to a set of geographical regions that form a partition on $\mathbf{D}$. Every node of $\mathcal{T}$ represents a region $r$, and subregions of $r$ are denoted by $r' \prec r$. The set of children $child(r) = \{r' \in \mathbf{R} | r' \prec r\}$ partitions the region $r$ in the next level of the tree. Each node has one parent region, indicated using $par(r)$. The tree has one root region called $r^{root}$. 

\paragraph{Queries.}
The dataset provider needs to release aggregate statistics on $\mathbf{D}$ using the set of $k$-way marginal queries $\mathbf{Q} = \{(q_1, q_2, \dots, q_k)\}$. A $k$-way marginal query consists of $k$ conditions, where each condition $q_k$ selects a particular value for the $k$\textsuperscript{th} feature from the corresponding domain $X_k$. The answer for a query counts the number of individuals whose features match the query predicates. Note that the features in a $k$-way query do not include geographic region, as we release the answers for all nodes of the region hierarchy tree for each query. We denote the set of ground truth answers for the query set $\mathbf{Q}$ on the dataset $\mathbf{D}$ using $\mathbf{A}_{\mathbf{Q}, \mathbf{D}}$.

To release aggregate statistics $\mathbf{A}_{\mathbf{Q}, \mathbf{D}}$ while protecting the privacy of individuals persons in $\mathbf{D}$ with formal guarantees, we use differential privacy \citep{dwork2014algorithmic}. Since differential privacy bounds the influence of any one person in the dataset on the output, the privatized answers will not be vulnerable to reconstruction attacks. We denote the set of privatized answers for the query set $\mathbf{Q}$ on the dataset $\mathbf{D}$ using $\tilde{\mathbf{A}}_{\mathbf{Q}, \mathbf{D}}$.

\paragraph{Hierarchical data constraints.} As well as being differentially private, we need $\tilde{\mathbf{A}}_{\mathbf{Q}, \mathbf{D}}$ to meet the following hierarchical data constraints \citep{fioretto2021differential} previously satisfied by $\mathbf{A}_{\mathbf{Q}, \mathbf{D}}$:
\begin{enumerate}
    \item Consistency: For each region $r \in \mathbf{R}$ and query $q \in \mathbf{Q}$, the answers in its subregions $r' \in child(r)$ add up to the answer for $r$, i.e., 
    $
        \tilde{a}_{q, \mathbf{D}}^{r} = \sum_{r' \in child(r)} \tilde{a}_{q, \mathbf{D}}^{r'}.
    $
    \item Validity: All answers $\tilde{a}_{q, \mathbf{D}}^{r}$ are non-negative integers. 
    \item Faithfulness: For each query $q$, the answers at each level $l$ of the region hierarchy tree add up to the same number, i.e., the true answer $G = a_{q, \mathbf{D}}^{r^{root}}$ for the root node, i.e.,
    $
        \sum_{r \in \mathbf{R}_l} \tilde{a}_{q, \mathbf{D}}^{r} = G.
    $
\end{enumerate}
Our goal is thus to release  $\tilde{\mathbf{A}}_{\mathbf{Q}, \mathbf{D}}$ solving the following optimization problem:

\begin{gather*}
    \min_{\tilde{\mathbf{A}}_{\mathbf{Q}, \mathbf{D}}}  \quad \|\tilde{\mathbf{A}}_{\mathbf{Q}, \mathbf{D}} - \mathbf{A}_{\mathbf{Q}, \mathbf{D}}\|_2\\
    \text{s.t.} \quad \tilde{a}_{q, \mathbf{D}}^{r} = \sum_{r' \in child(r)} \tilde{a}_{q, \mathbf{D}}^{r'}; \forall q \ \forall r \quad \tilde{a}_{q,\mathbf{D}}^{r} \in \mathbf{Z}^+;\\
    \forall q \ \forall l \quad \sum_{r \in \mathbf{R}_l} \tilde{a}_{q, \mathbf{D}}^{r} = G
\end{gather*}

\subsection{Algorithms}
\label{sec:alg-desc}
We consider two classes of algorithms: DP release of hierarchical statistics, and DP synthetic data. 

\paragraph{Private statistic release.}
The TopDown algorithm is the best-known algorithm for releasing hierarchical tabular statistics \citep{abowd20222020}. 
It consists of two steps. First, random noise is added to the ground truth aggregate statistics $\mathbf{A}_{\mathbf{Q}, \mathbf{D}}$ to create $\hat{\mathbf{A}}_{\mathbf{Q}, \mathbf{D}}$. This noise is sampled from a double-geometric distribution with a parameter of $\frac{2L}{\epsilon}$, where $L$ is the number of levels in the region hierarchy tree. Next, $\hat{\mathbf{A}}_{\mathbf{Q}, \mathbf{D}}$ is post-processed using an optimization routine to satisfy the hierarchical data constraints. Note that to run the TopDown algorithm, we need to know the full set of queries ahead of time, and it is not possible to submit out-of-distribution queries post-release. In our experiments, we use the optimized version of \citet{fioretto2021differential}, which provides state-of-the-art utility for this class of methods.
The TopDown algorithm satisfies $\epsilon$-differential privacy. 

\paragraph{Private synthetic data.}
When generating private synthetic data, answers to the query set $\mathbf{Q}$ are computed on a synthetic dataset $\mathbf{D'}$ instead of adding noise to answers computed using the actual dataset $\mathbf{D}$. This means that we release $\mathbf{A}_{\mathbf{Q}, \mathbf{D'}}$ instead of $\Tilde{\mathbf{A}}_{\mathbf{Q}, \mathbf{D}}$, and the hierarchical data constraints are already met. 
We consider two synthetic data algorithms in this work: Hierarchical Product Distributions (HPD) \cite{liu2022private} and Maximum Spanning Tree (MST) \cite{mckenna2021winning}.

\textit{Hierarchical Product Distributions (HPD):}
To our knowledge, HPD is the only private synthetic data method that is explicitly designed to model hierarchical data distributions.  
Since census data consists of features with large domain sizes, \citet{liu2022private} propose modeling the underlying data distribution as a mixture of hierarchical product distributions (HPD) over group-level (e.g., county) and individual-level features (e.g., age). The iterative Adaptive Measurements framework is then used to solve for $\mathbf{D'}$. HPDs make solving for $\mathbf{D'}$ more tractable.

A learned HPD is utilized in the following manner. Firstly, a group-level feature $m$ and the number of individuals $n_m$ in that group are randomly sampled. Subsequently, the individual-level features for people in this group are sampled $n_m$ times. The authors consider two variants based on how the HPD is parameterized: HPD-Fixed and HPD-Gen \citep{liu2022private}. HPD-Fixed defines a fixed distribution whose parameters are the exact probability vectors in the product distributions. HPD-Gen uses a generative neural network that learns from the original data and creates samples from random Gaussian noise according to the learned distribution. 

In each round of the Adaptive Measurements framework, the following three steps are performed. First, a query $q$ that has high error is sampled privately from $\mathbf{Q}$ using the exponential mechanism. Next, the answer to $q$ is privately computed on $\mathbf{D}$ using the Gaussian mechanism. Finally, an optimization problem solves a loss function that minimizes the error of $\mathbf{D'}$ on all queries, including the current round. It is worth noting that HPD-Fixed and HPD-Gen satisfy $(\epsilon, \delta)$-differential privacy.

\textit{Maximum Spanning Tree (MST):}
Although the MST synthetic data algorithm is not designed for hierarchical data distributions, it is the state-of-the-art tabular synthetic data algorithm,  achieving the best results in the recent work of \citet{tao2021benchmarking} on benchmarking private synthetic data algorithms, where it outperformed a host of other synthetic data algorithms \citep{aydore2021differentially, chanyaswad2017ron, ge2020kamino, gretelrnn2023, jordon2018pate, mckenna2019graphical, rosenblatt2020differentially, vietri2020new, xie2018differentially, zhang2017privbayes}.
Instead of using $\mathbf{Q}$, the MST algorithm selects a query set $\mathbf{Q'}$ from all possible $2-, \dots, k$-way queries. $\mathbf{Q'}$ is constructed by selecting features that form a maximum spanning tree of a correlation graph, where the nodes correspond to features and edges correspond to the mutual information between two features \citep{mckenna2021winning}. Since the correlation graph is based on the actual dataset $\mathbf{D}$, the exponential mechanism is used to select features. This selection process consumes part of the privacy budget. The queries in $\mathbf{Q'}$ are privately measured using the Gaussian mechanism. The Private-PGM post-processing method uses these measurements to learn a high-dimensional data distribution \citep{mckenna2019graphical}, which is used to generate synthetic data. Like HPD-Fixed and HPD-Gen, MST satisfies $(\epsilon, \delta)$-differential privacy.

\begin{figure*}[ht!]
    \centering
    \begin{subfigure}[b]{0.48\textwidth}
    \centering
    \includegraphics[width=\textwidth]{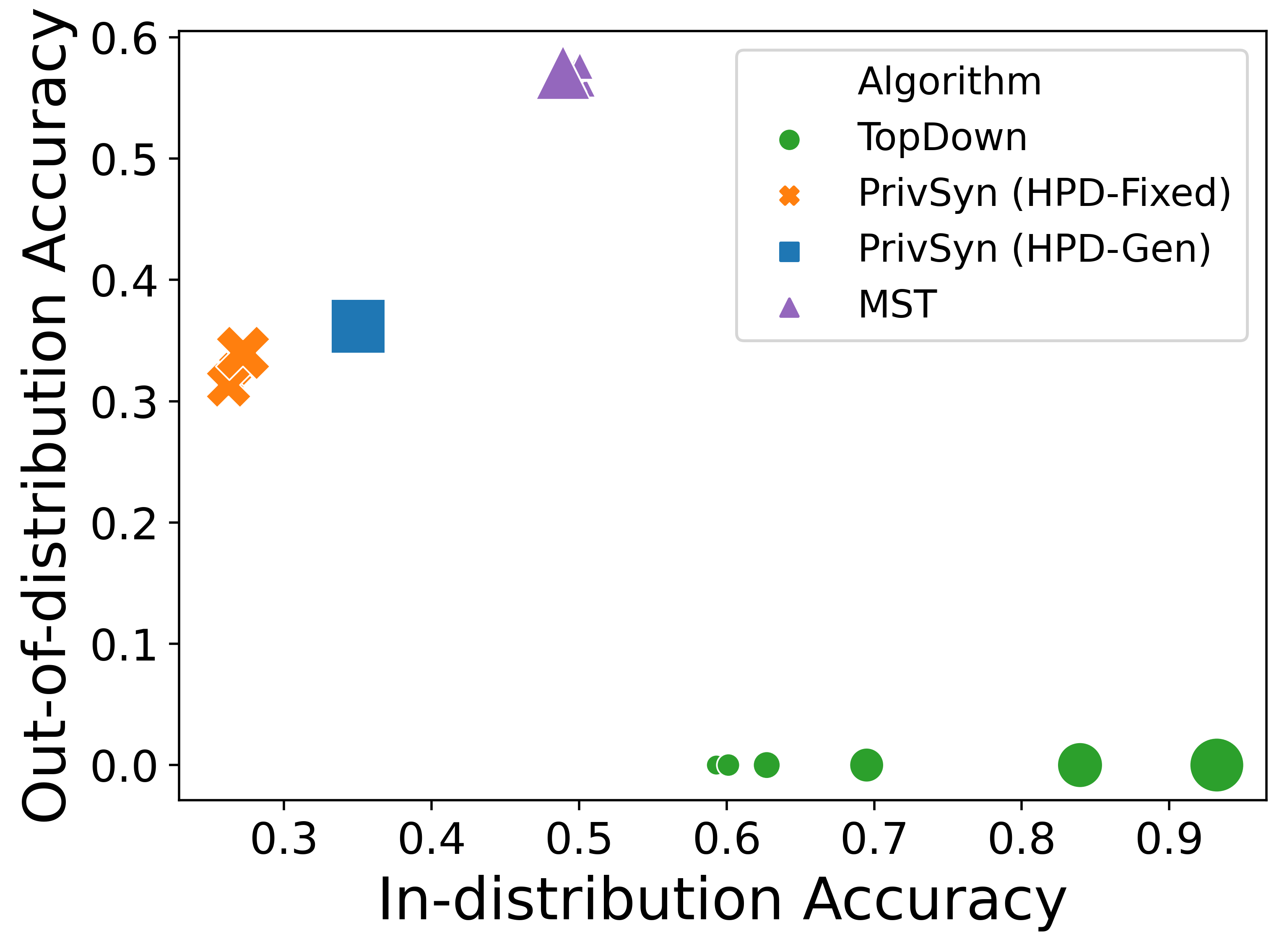}
    \caption{In- vs. out-of-distribution accuracy (NY19).}
    \label{fig:avg-acc-3-way-queries}
    \end{subfigure}
    \hfill
    \begin{subfigure}[b]{0.48\textwidth}
    \centering
    \includegraphics[width=\textwidth]{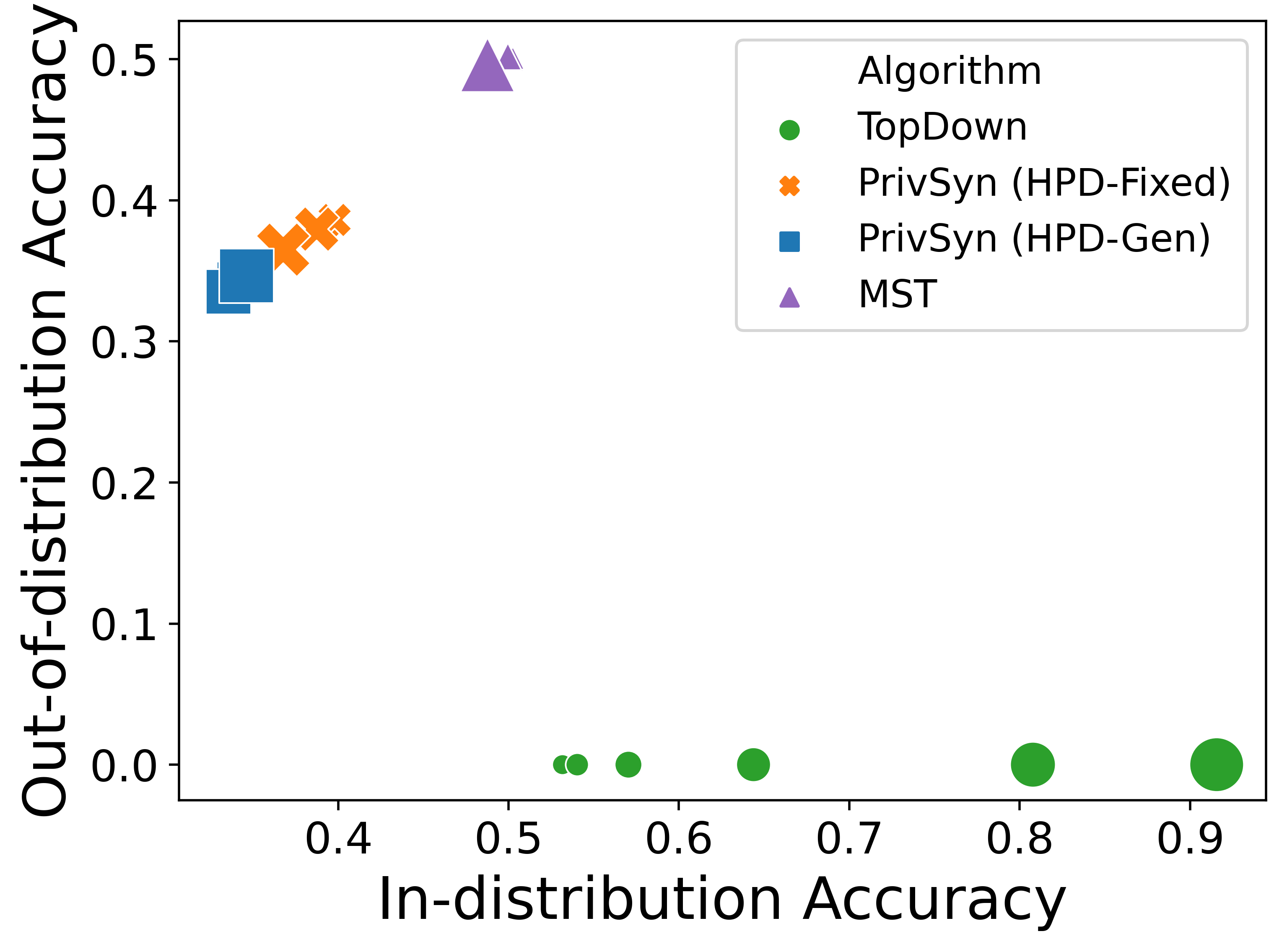}
    \caption{In- vs. out-of-distribution accuracy (US21).}
    \label{fig:pb-avg-acc-3-way-queries}
    \end{subfigure}
    \caption{Accuracy of algorithms on in-distribution and out-of-distribution three-way marginal queries on ACS NY 2019 (left) and ACS Public Coverage 2021 (right). Accuracy describes the fraction of queries whose answers match ground truth. Both HPD methods are trained on queries used in the evaluation, whereas MST does not use prior knowledge of these queries. The size of the markers increases with  $\epsilon \in \{0.125, 0.25, 0.5, 1, 2, 3\}$. All metrics are averaged over 20 runs.
    }
    \label{fig:ny-pb-acc-3-way-queries}
\end{figure*}

\section{Experiments}
\paragraph{Datasets.} We ran our evaluation on two demographic datasets, one at the regional level (New York state) and the other at the national level (the US at large). The American Community Survey (ACS) dataset contains information about the United States' changing population, housing, and workforce. For the first dataset (ACS NY 2019, abbreviated NY19), we modify the data as suggested in \citep{liu2022private} by selecting households of size ten or smaller from the state of New York in 2019 and discretizing all 21 features. ACS NY 2019 contains records of $n = 197512$ individuals, and the regional hierarchy tree has two levels - county and state - with COUNTYFIP encoding an individual's county. For the second dataset (ACS Public Coverage 2021, abbreviated US21), we follow the filtering procedure in \citep{ding2021retiring}, select up to 10,000 records from each state in the US, and discretize all 14 features. ACS Public Coverage 2021 contains records of $n = 212900$ individuals, and the regional hierarchy tree has two levels - state and country - with ST encoding an individual's state. Although we were able to find some publicly-available real datasets from non-US contexts \citep{kaggle_international_data, eurostat_population_housing_censuses, census_india_tables, singstat_tablebuilder}, they are all pre-aggregated, and hence cannot be used to run our experiments.

\paragraph{Implementation details.} For the TopDown algorithm, we used an implementation provided by \citet{fioretto2021differential}. Our only modifications to the code were to adapt it to use the ACS datasets and the same in-distribution queries for generating the synthetic datasets. For HPD-Fixed and HPD-Gen, we used an implementation provided by \citet{liu2022private}. We used the default hyperparameters; HPD-Fixed uses a learning rate of 0.1. HPD-Gen uses a learning rate of 0.0001 and has 512 hidden layers in its generative neural network. Both HPD-Fixed and HPD-Gen are run for 100 iterations. We used a fixed $\delta = 1/n^2$ for both variants. For MST, we used the implementation \citep{ryan112358_private_pgm_mst} released by \citet{mckenna2019graphical} and set $\delta = 1/n^2$.

\paragraph{Query complexity and privacy.} We assess how well private tabular data release methods perform for queries with complexity $k \in \{1, 2, 3\}$. In particular, we use 137 one-way, 1295 two-way, and 3196 three-way marginal queries, selected uniformly at random for ACS NY 2019 (we generate similar number of queries for ACS Public Coverage 2021). To generate synthetic datasets, we train the PrivSyn algorithms on the exact marginal queries used in the evaluation. These queries are called in-distribution queries. The MST algorithm does not use prior knowledge of queries during training. 
We also select out-of-distribution queries for evaluation, and evaluate all three baselines on both in- and out-of-distribution queries (though this distinction is not meaningful for MST). We repeat the evaluation for various $\epsilon \in \{0.125, 0.25, 0.5, 1, 2, 3\}$.

\subsection{Metrics}
\label{sec:metrics}
\paragraph{Comparing private statistics to synthetic data.} For each query, we compute the answers for all nodes in the regional hierarchy tree (i.e., all counties, the state total). Since the queries are all numeric, we plot the distribution of absolute errors with respect to the ground truth dataset for each algorithm, i.e., $\forall q \in \mathbf{Q}$, we plot $|\tilde{a}_{q, D} - {a}_{q, D}|$ for the TopDown algorithm and $|a_{q, D'} - {a}_{q, D}|$ for the private synthetic datasets. To compare errors of two algorithms, we plot the distribution of the difference between their absolute errors for various $k$ and $\epsilon$, i.e., $\forall q \in \mathbf{Q}$, we plot $|\tilde{a}_{q, D} - {a}_{q, D'}|$. Since the TopDown algorithm cannot answer out-of-distribution queries, we only compare absolute errors for in-distribution queries. We also calculate the accuracy of each algorithm as the fraction of queries it answers correctly, i.e., the algorithm's answer matches the ground truth dataset's answer. On out-of-distribution queries, we consider TopDown to have 0 accuracy. 

\begin{figure*}[ht!]
    \centering
    \begin{subfigure}[b]{0.48\textwidth}
    \includegraphics[width=\textwidth]{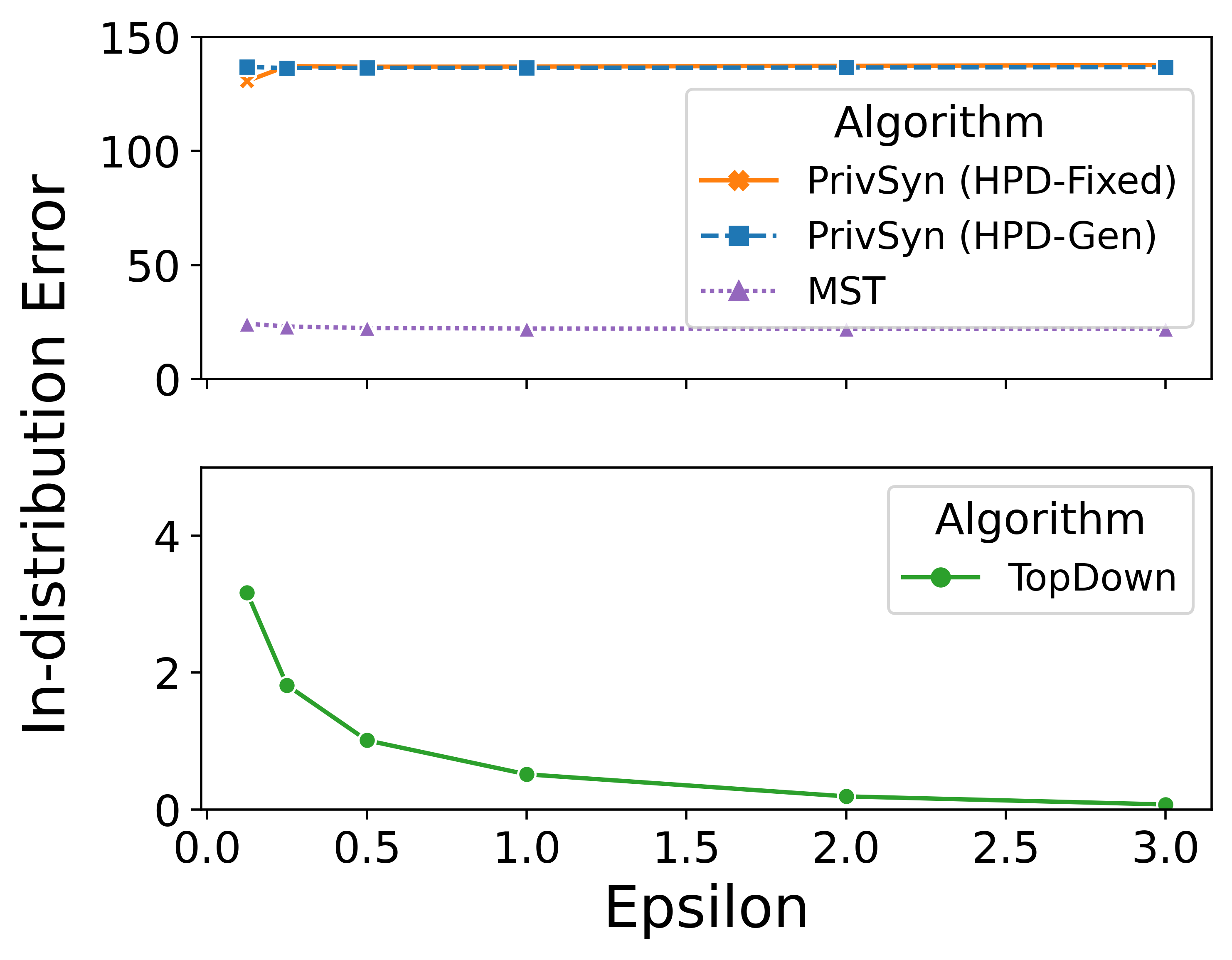}
    \caption{Average absolute error, in-distribution (NY19).}
    \label{fig:in-dist-err-3-way-queries}
    \end{subfigure}
    \hfill
    \begin{subfigure}[b]{0.48\textwidth}
    \includegraphics[width=\textwidth]{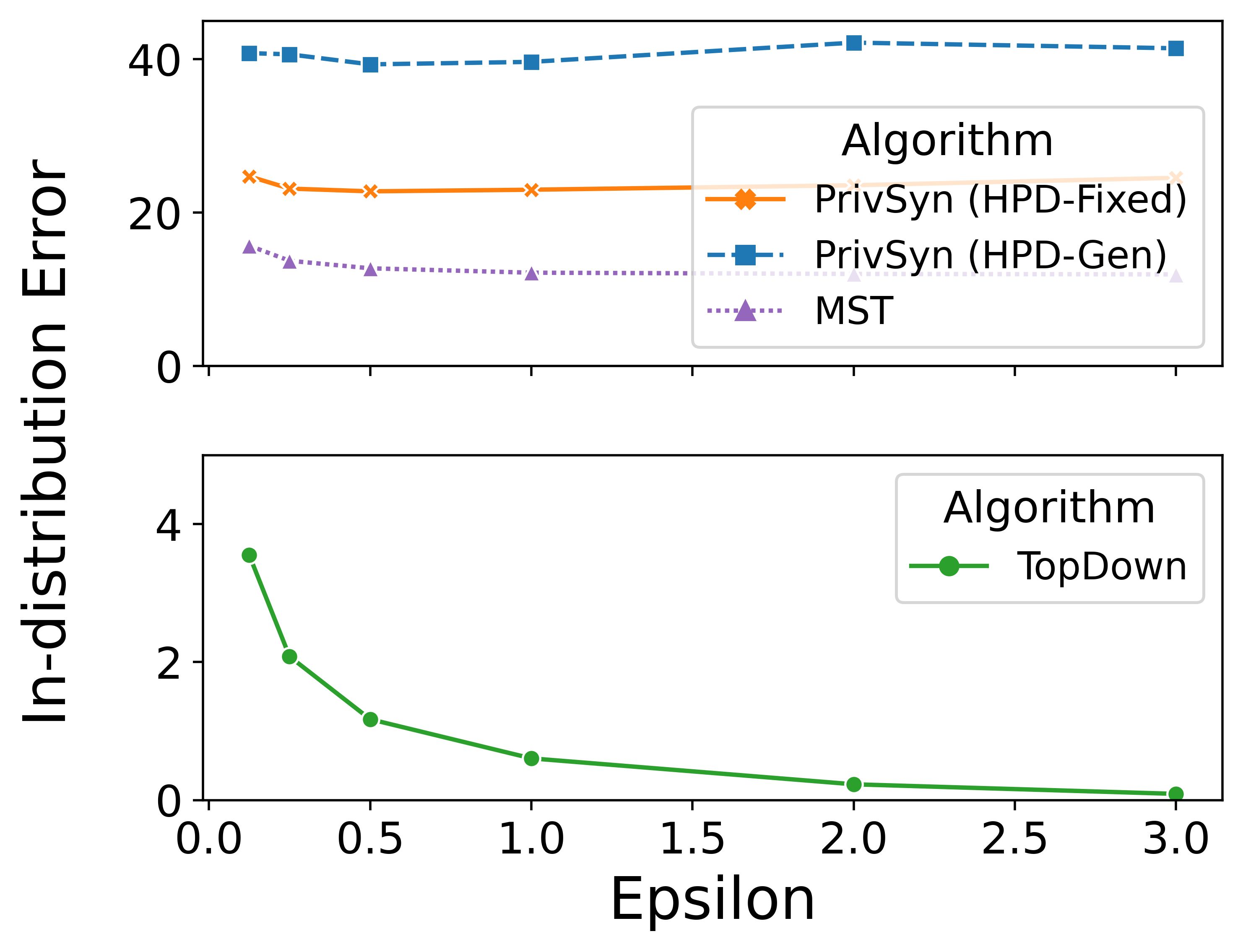}
    \caption{Average absolute error, in-distribution (US21).}
    \label{fig:pb-in-dist-err-3-way-queries}
    \end{subfigure}
    \caption{Average absolute error of algorithms on in-distribution three-way marginal queries on ACS NY 2019 (left) and ACS Public Coverage 2021 (right). Absolute error between the answers returned by the algorithm and those on the ground truth dataset are averaged over 20 runs over all in-distribution queries since the TopDown algorithm cannot answer out-of-distribution queries. Both HPD methods are trained on queries used in the evaluation, whereas MST does not use prior knowledge of these queries. 
    }
    \label{fig:eps-vs-in-dist-err}
\end{figure*}

\paragraph{Evaluating synthetic data quality.} We compare the synthetic datasets produced by the HPD-Fixed, HPD-Gen, and MST algorithms using the metrics introduced in \citep{tao2021benchmarking}. In particular, we compute Individual Attribute Distribution Similarity (\textbf{Ind}), Pairwise Attribute Distribution Similarity (\textbf{Pair}), and Pairwise Correlation Similarity (\textbf{Corr}) at different values of $\epsilon \in \{0.125, 0.25, 0.5, 1, 2, 3\}$. Ind is the average total variation distance (TVD) of all one-way marginal distributions (i.e., $M$ features) with respect to the ground truth dataset. Similarly, Pair is the average TVD of all pairs of two-way marginal distributions. Corr measures how many pairs of features have the same correlation level among the synthetic and ground truth datasets. We compute the Cramer's V between each pair of features and bucket it into four levels as suggested in \citep{tao2021benchmarking}: $V \in [0, .1)$ = ``low", $[.1, .3)$ is ``weak", $[.3, .5)$ is ``middle" and $[.5, 1)$ is ``strong''. 

\subsection{Results}
The performance of the two approaches on in-distribution and out-of-distribution three-way marginal queries is depicted in Figures \ref{fig:ny-pb-acc-3-way-queries} and \ref{fig:eps-vs-in-dist-err}. 
Figure \ref{fig:ny-pb-acc-3-way-queries} shows the tradeoff between in-distribution and out-of-distribution accuracy for both datasets. 
Since TopDown cannot answer out-of-distribution queries, we consider it to have 0 accuracy on these queries.
As expected, synthetic data does much better on out-of-distribution queries, whereas TopDown does much better on in-distribution queries. 
However, the scale of difference on in-distribution queries is notable.
At $\epsilon=0.125$, both TopDown and MST achieve similar in-distribution accuracy. However, at a more realistic privacy budget of $\epsilon=3$, the TopDown algorithm has an accuracy of 0.92 compared to 0.49 for the best performing synthetic data algorithm (MST) -- a difference of $1.9\times$ at the same $\epsilon$.

\begin{figure*}[ht!]
     \centering
     \begin{subfigure}[b]{0.32\textwidth}
         \centering\includegraphics[width=\textwidth]{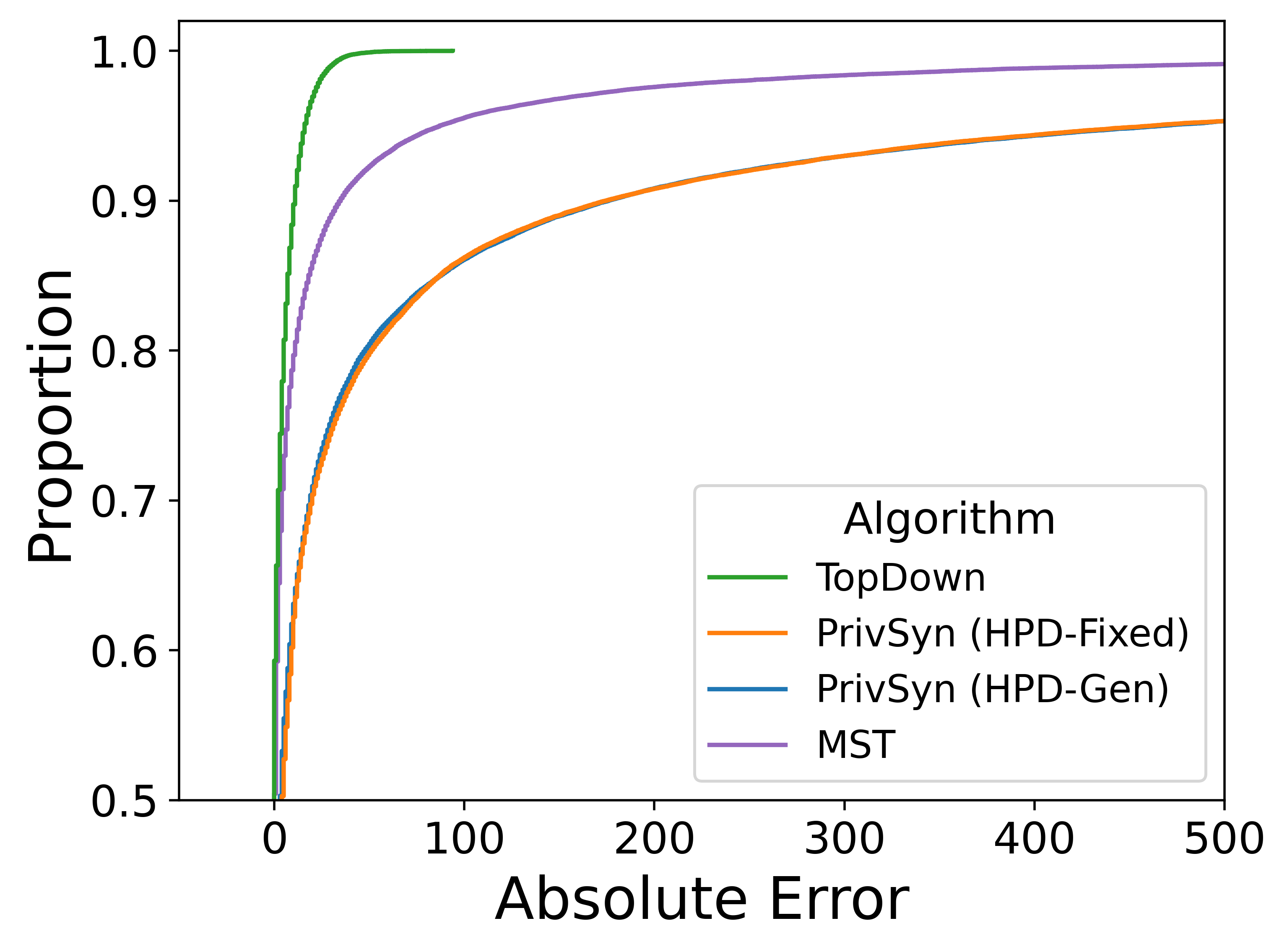}
         \caption{$\epsilon = 0.125$}
         \label{fig:eps1}
     \end{subfigure}
     \hfill
     \begin{subfigure}[b]{0.32\textwidth}
         \centering
         \includegraphics[width=\textwidth]{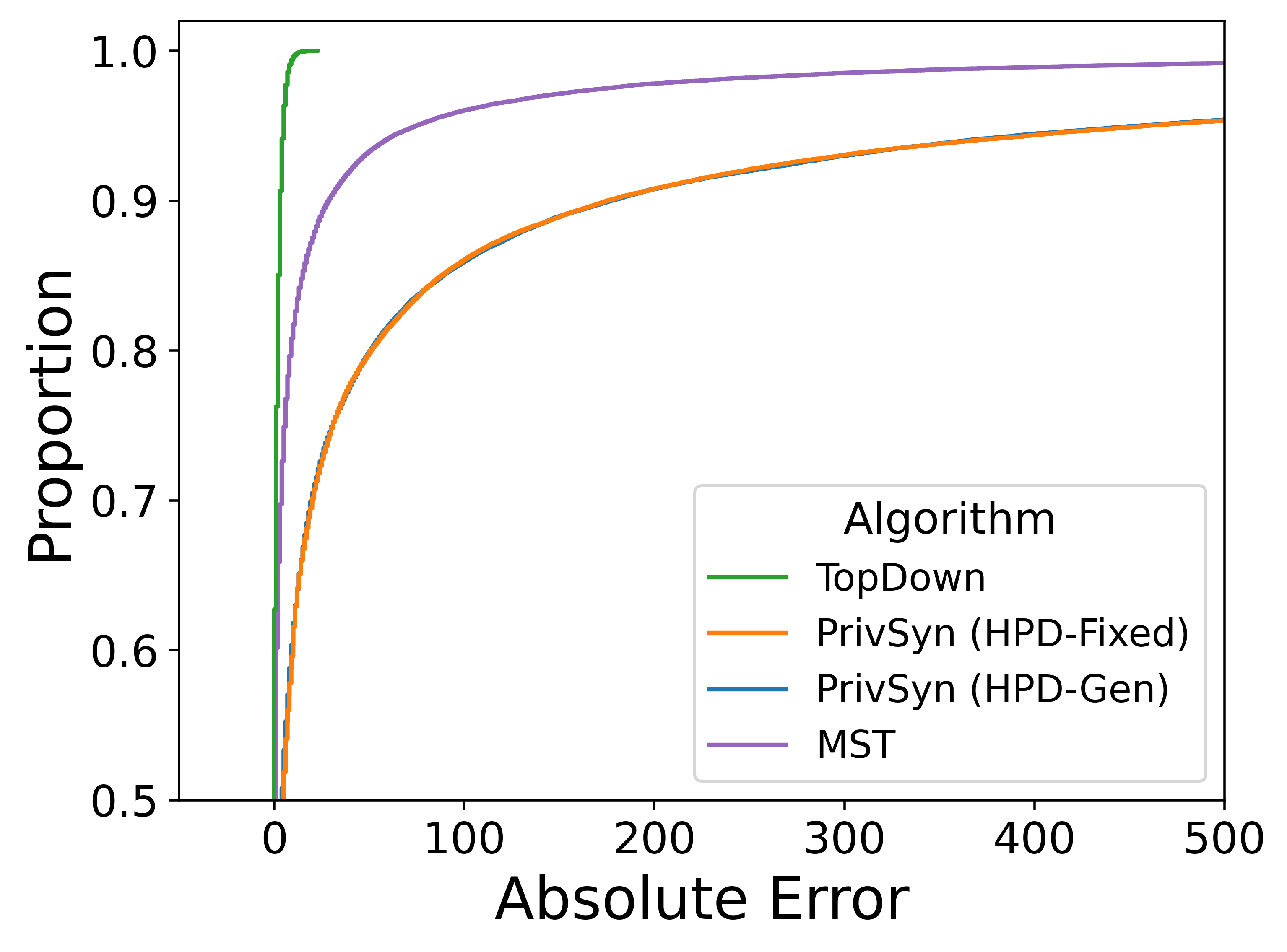}
         \caption{$\epsilon = 0.5$}
         \label{fig:eps2}
     \end{subfigure}
     \hfill
     \begin{subfigure}[b]{0.32\textwidth}
         \centering
         \includegraphics[width=\textwidth]{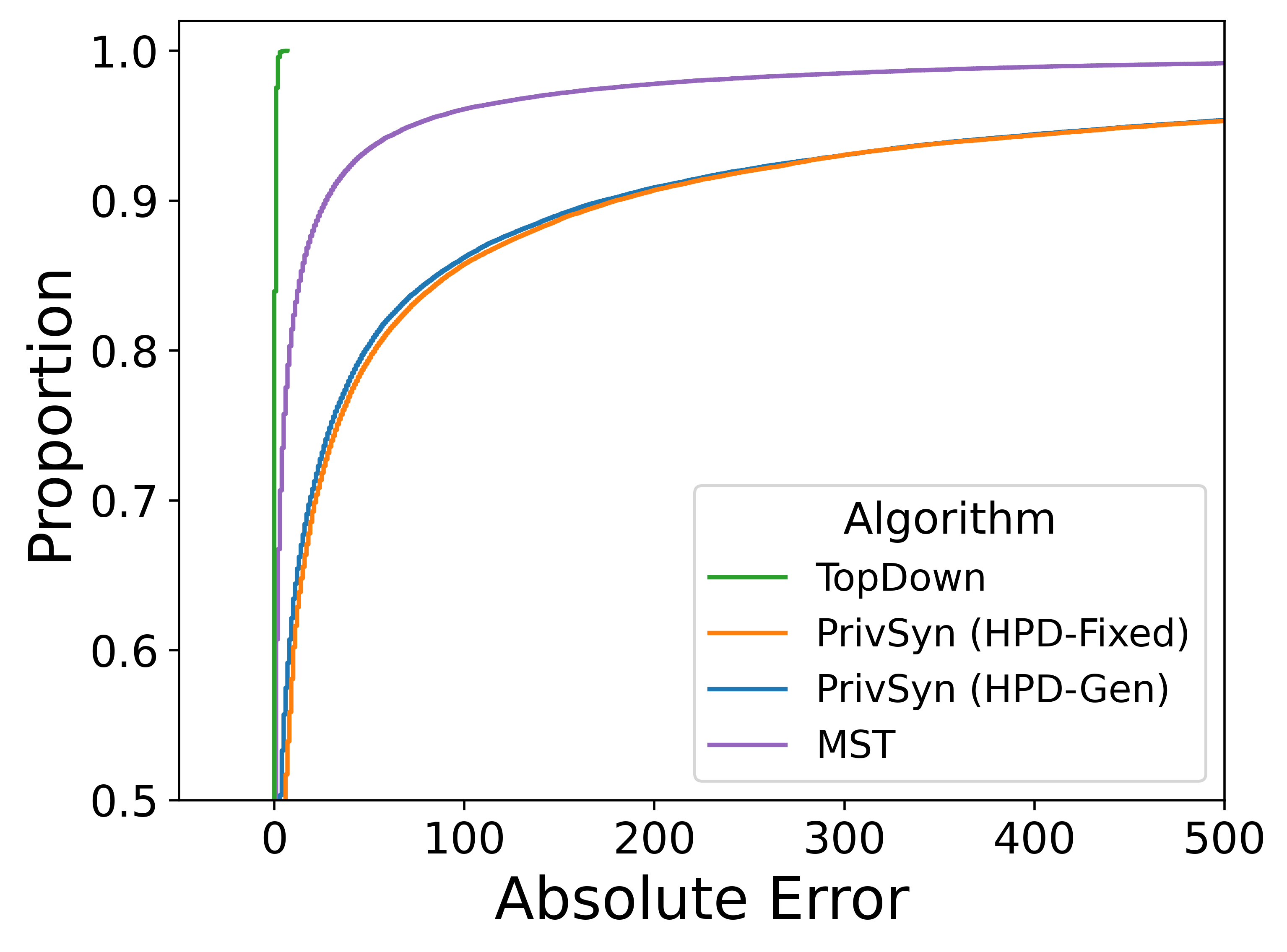}
         \caption{$\epsilon = 2.0$}
         \label{fig:eps3}
     \end{subfigure}
        \caption{Absolute error CDFs for TopDown, HPD-Fixed, HPD-Gen, and MST for in-distribution three-way marginal queries ($k = 3$) on ACS NY 2019. As $\epsilon$ increases (reduced privacy, higher accuracy), the error distributions of TopDown skew towards 0.}
        \label{fig:abs-error}
\end{figure*}

For a more nuanced look, Figure \ref{fig:eps-vs-in-dist-err} shows the effect of $\epsilon$ on in-distribution average absolute error (the Top-Down algorithm cannot compute out-of-distribution queries).
As expected, the TopDown algorithm's absolute error decreases as $\epsilon$ increases (reduced privacy).  It also outperforms all private synthetic datasets on in-distribution queries at all $\epsilon$ values by a significant margin (at least $4\times$). Figure \ref{fig:abs-error} further shows the full absolute error CDF at fixed values of $\epsilon$. 

Note in Figure \ref{fig:eps-vs-in-dist-err} that for private synthetic datasets, the average error does not always decrease as $\epsilon$ increases. One possible explanation for this is given by \citet{alvim2023novel}. Private synthetic datasets use ``geometric perturbers'', e.g., the Gaussian mechanism, to introduce noise. \citet{alvim2023novel} show that such perturbers can be unstable with respect to utility (e.g., average absolute error) computed over ``post-processors''---that is, the set of marginal queries used in our evaluation. An unstable method does not preserve the increase in $\epsilon$-decrease in utility trend after post-processing. 
We hypothesize that their observations may help explain the non-monotonicity in Figure \ref{fig:eps-vs-in-dist-err}.

Another detail to note in our results is the following: 
as described in \S\ref{sec:alg-desc}, the optimized version of the TopDown algorithm we evaluate \cite{fioretto2021differential} satisfies $\epsilon$-differential privacy, whereas HPD-Fixed, HPD-Gen, and MST satisfy $(\epsilon, \delta)$-differential privacy. 
Unlike the optimized version, the original TopDown algorithm \citep{abowd20222020} satisfies zero-concentrated differential privacy (zCDP) \citep{bun2016concentrated} and uses a discrete Gaussian mechanism to introduce noise. 
We might observe a larger gap in utility between TopDown and private synthetic data if we modify the optimized TopDown algorithm to use the discrete Gaussian mechanism instead of the Geometric mechanism. 
Furthermore, using zCDP would allow us to translate TopDown's privacy guarantee to $(\epsilon, \delta)$-differential privacy, giving us a better comparison with the private synthetic data algorithms. 
We benchmarked only existing methods in our study, and did not do this comparison; however, extending the TopDown algorithm to use zCDP and evaluate its performance would be an interesting direction for future work.

\begin{figure*}[ht!]
     \centering
     \begin{subfigure}[b]{0.329\textwidth}
         \centering
         \includegraphics[width=\textwidth]{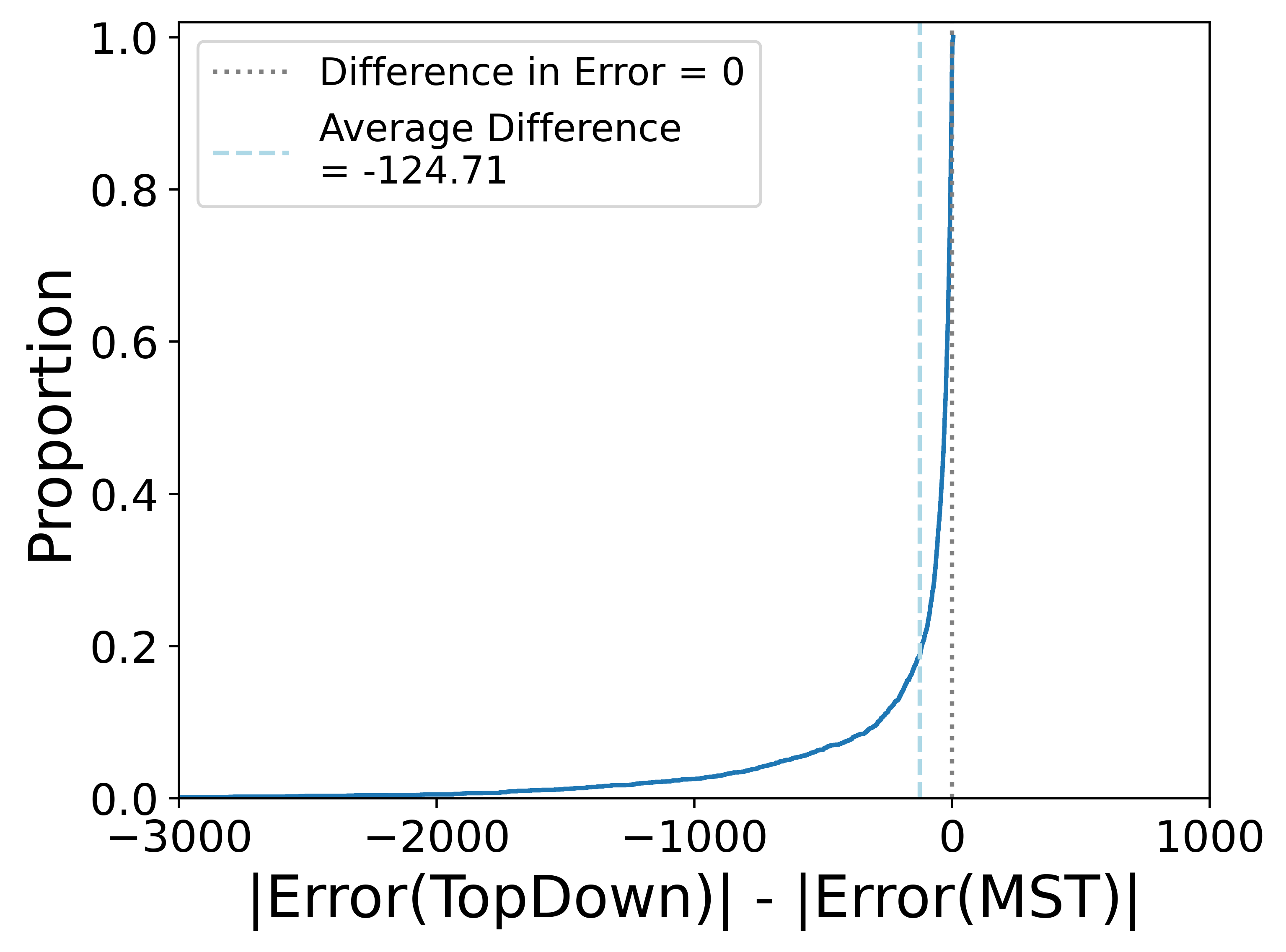}
         \caption{\textbf{One}-way marginal queries.}
         \label{fig:oneway}
     \end{subfigure}
     \hfill
     \begin{subfigure}[b]{0.329\textwidth}
         \centering
         \includegraphics[width=\textwidth]{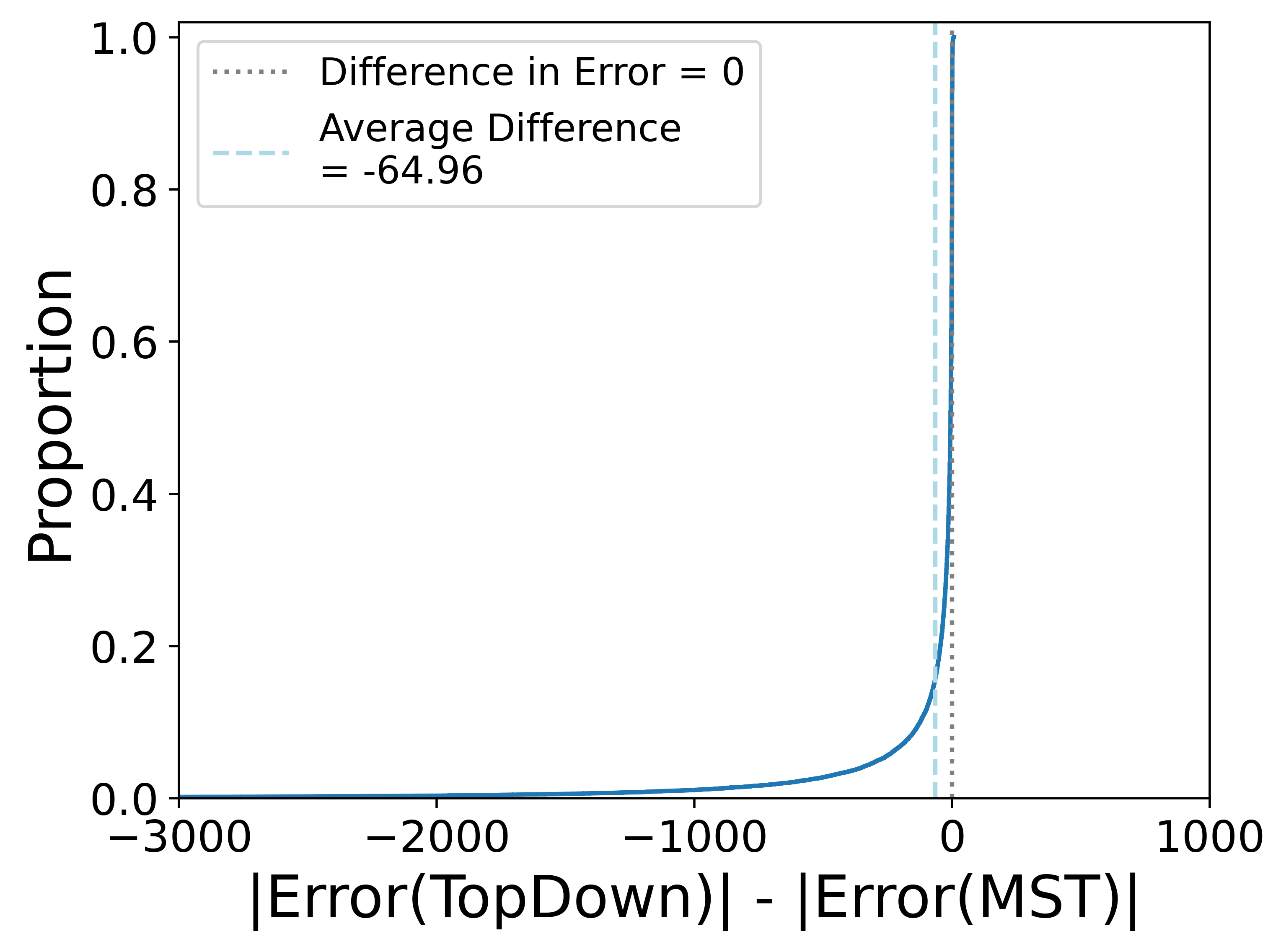}
         \caption{\textbf{Two}-way marginal queries.}
         \label{fig:twoway}
     \end{subfigure}
     \hfill
     \begin{subfigure}[b]{0.329\textwidth}
         \centering
         \includegraphics[width=\textwidth]{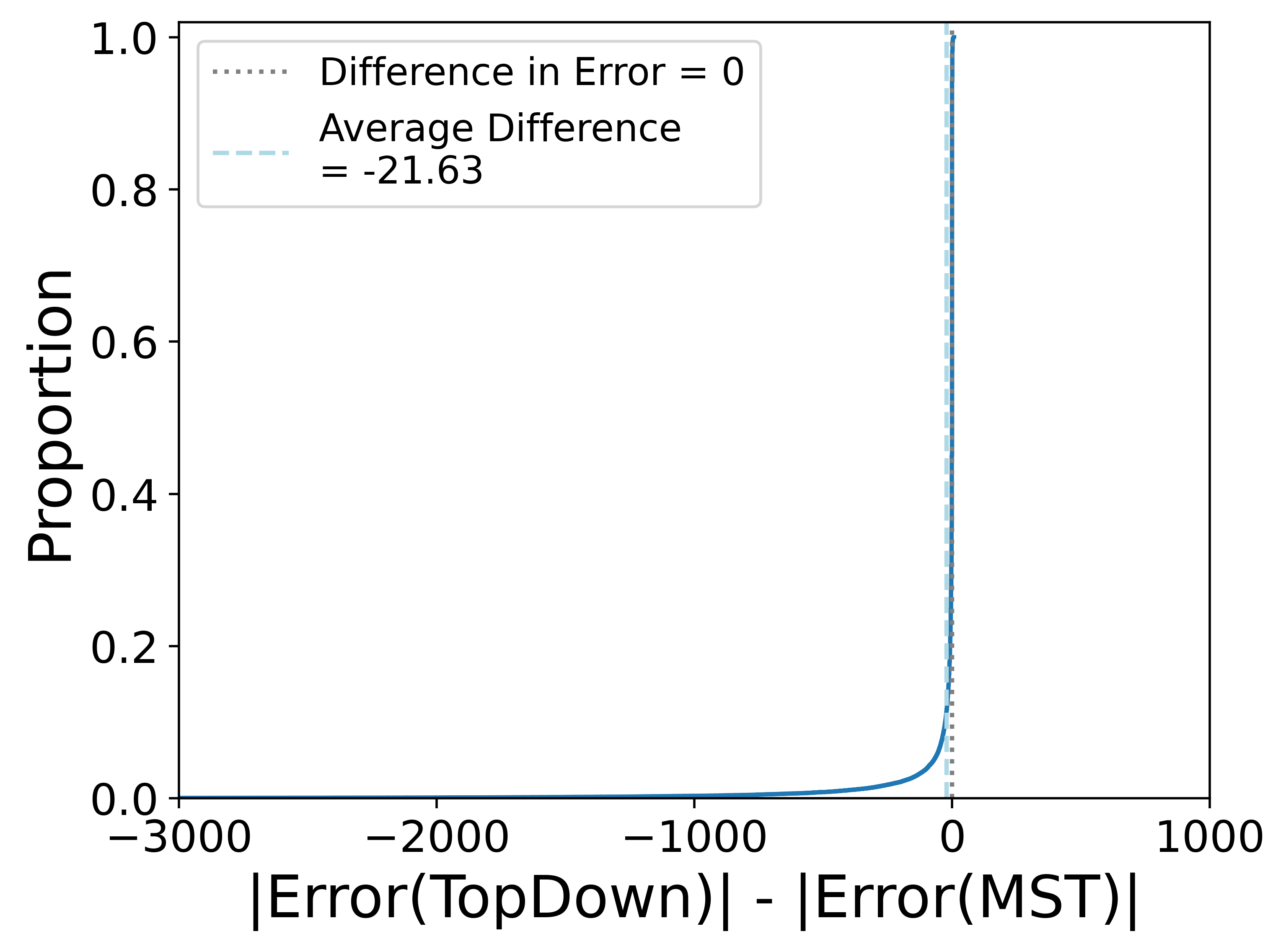}
         \caption{\textbf{Three}-way marginal queries.}
         \label{fig:threeway}
     \end{subfigure}
        \caption{Difference between absolute errors for TopDown and MST for each query at $\epsilon = 1.0$ on ACS NY 2019. Positive values indicate TopDown incurs higher absolute error, whereas negative values indicate the private synthetic dataset incurs higher error. As the complexity of the query increases, the error of the synthetic dataset approaches that of the TopDown algorithm.}
        \label{fig:diff-bw-abs-error}
\end{figure*}

We next study the effect of query complexity on performance. In Figure \ref{fig:diff-bw-abs-error}, we show the CDF of the difference between the absolute errors of TopDown and MST for $k \in \{1, 2, 3\}$ marginal in-distribution queries. The TopDown algorithm performs better than MST for all queries. However, as the query complexity increases, the performance of the private synthetic dataset improves and becomes comparable to the TopDown algorithm. These results suggest that synthetic data may be a good option when mostly complex marginal queries (i.e., $k\geq 3$) need to be made.

\begin{figure*}[ht!]
     \centering
     \begin{subfigure}[b]{0.329\textwidth}
         \centering
         \includegraphics[width=\textwidth]{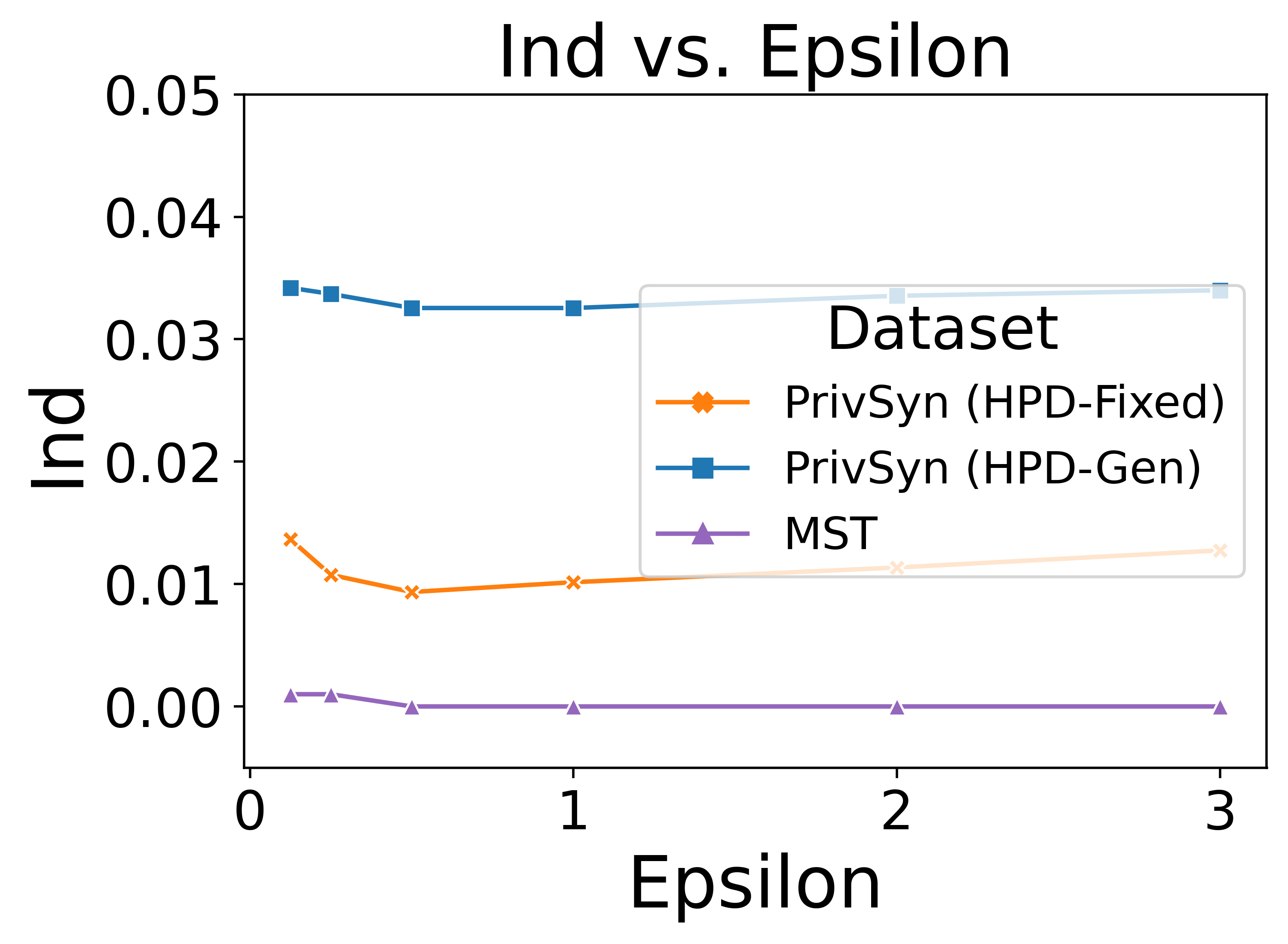}
         \caption{Individual Attribute Distribution Similarity (\textbf{Ind}) v.s. $\epsilon$. Lower values are better.}
         \label{fig:ind}
     \end{subfigure}
     \hfill
     \begin{subfigure}[b]{0.329\textwidth}
         \centering
         \includegraphics[width=\textwidth]{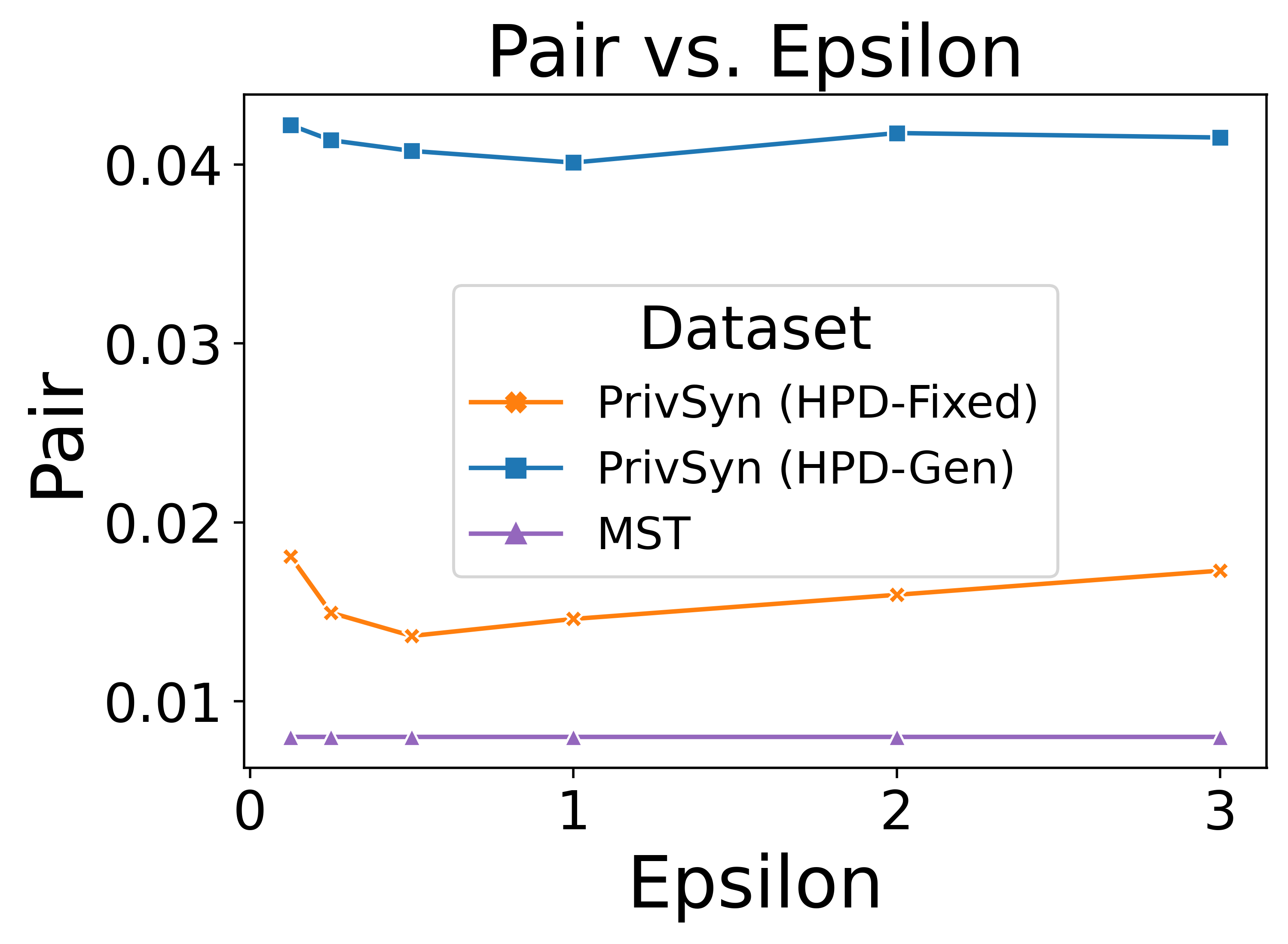}
         \caption{Pairwise Attribute Distribution Similarity (\textbf{Pair}) v.s. $\epsilon$. Lower values are better.}
         \label{fig:pair}
     \end{subfigure}
     \hfill
     \begin{subfigure}[b]{0.329\textwidth}
         \centering
         \includegraphics[width=\textwidth]{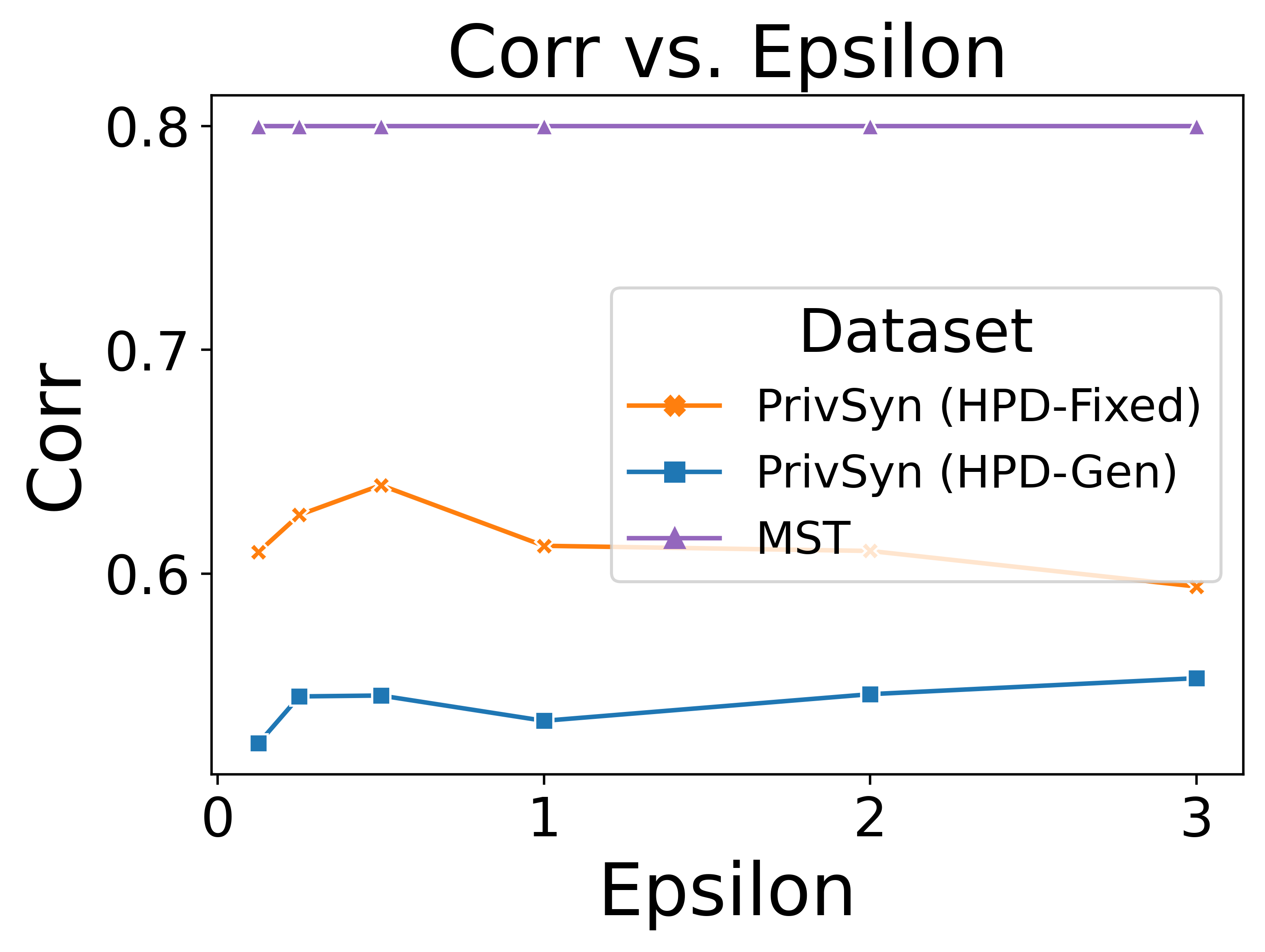}
         \caption{Pairwise Correlation Similarity (\textbf{Corr}) v.s. $\epsilon$. Higher values are better.}
         \label{fig:corr}
     \end{subfigure}
        \caption{Performance metrics \citep{tao2021benchmarking} for synthetic data algorithms trained on 3-way marginal queries at $\epsilon \in \{0.125, 0.25, 0.5, 1, 2, 3\}$ on ACS Public Coverage 2021.}
        \label{fig:pb-syn-benchmark}
\end{figure*}

\paragraph{Synthetic data detailed comparison.}
Finally, in Figure \ref{fig:pb-syn-benchmark}, we assess the three synthetic datasets' ability to approximate the actual data distribution at different $\epsilon$ values for ACS Public Coverage 2021 (we observed similar trends for ACS NY 2019). 
We used metrics drawn from the benchmarking work of \citet{tao2021benchmarking}. 
As defined in \S\ref{sec:metrics}, the Ind and Pair metrics measure the Total Variation Distance between the distributions of individual features and pairs of features, respectively (lower is better). Corr measures how many pairs of features have the same correlation level (Cramer's V) for pairs of features, between the ground truth and synthetic dataset. MST better preserves the individual feature distributions, as well as distributions between pairs of features and the correlation between them. This is likely because MST selects better queries to learn the underlying data distribution \citep{mckenna2021winning}, whereas HPD-Fixed and HPD-Gen use the queries we randomly generated for the evaluation. 
Understanding exactly why MST performs better than the HPD methods is a question left for future work. 

\section{Discussion}

Our results suggest that if queries are known ahead of time, the TopDown algorithm is (for now) the clear winner, for all query complexities. 
However, private synthetic data generation algorithms should be used when dataset providers want to release answers to queries that are not known beforehand, or if the data needs to be shared in its original format. 
Our results suggest several important research directions, including methods for improving DP synthetic data algorithms over hierarchical datasets, and dynamically updating the TopDown method when the queries are unknown \emph{a priori}. 
Moreover, it is currently unclear how to use either method in a dynamic setting, i.e., when data users want to make repeated queries that adapt to the population changing over time.
We believe these pose interesting and important challenges for the DP and synthetic data research communities.

\section*{Acknowledgments}
The authors gratefully acknowledge the Air Force Office of Scientific Research under award number FA9550-21-1-0090, NSF grant CNS-2148359, the Bill \& Melinda Gates Foundation, Intel, and the Sloan Foundation for their generous support.

\bibliography{aaai24}

\begin{thebibliography}{30}
\providecommand{\natexlab}[1]{#1}

\bibitem[{Abowd et~al.(2022)Abowd, Ashmead, Cumings-Menon, Garfinkel, Heineck, Heiss, Johns, Kifer, Leclerc, Machanavajjhala, and Moran}]{abowd20222020}
Abowd, J.~M.; Ashmead, R.; Cumings-Menon, R.; Garfinkel, S.; Heineck, M.; Heiss, C.; Johns, R.; Kifer, D.; Leclerc, P.; Machanavajjhala, A.; and Moran, B. 2022.
\newblock The 2020 census disclosure avoidance system topdown algorithm.
\newblock \emph{Harvard Data Science Review}, (Special Issue 2).

\bibitem[{Alvim et~al.(2023)Alvim, Fernandes, McIver, Morgan, and Nunes}]{alvim2023novel}
Alvim, M.~S.; Fernandes, N.; McIver, A.; Morgan, C.; and Nunes, G.~H. 2023.
\newblock A novel analysis of utility in privacy pipelines, using Kronecker products and quantitative information flow.
\newblock In \emph{Proceedings of the 2023 ACM SIGSAC Conference on Computer and Communications Security}, 1718--1731.

\bibitem[{Aydore et~al.(2021)Aydore, Brown, Kearns, Kenthapadi, Melis, Roth, and Siva}]{aydore2021differentially}
Aydore, S.; Brown, W.; Kearns, M.; Kenthapadi, K.; Melis, L.; Roth, A.; and Siva, A.~A. 2021.
\newblock Differentially private query release through adaptive projection.
\newblock In \emph{International Conference on Machine Learning}, 457--467. PMLR.

\bibitem[{Bun and Steinke(2016)}]{bun2016concentrated}
Bun, M.; and Steinke, T. 2016.
\newblock Concentrated differential privacy: Simplifications, extensions, and lower bounds.
\newblock In \emph{Theory of Cryptography Conference}, 635--658. Springer.

\bibitem[{{Census of India}(2023)}]{census_india_tables}
{Census of India}. 2023.
\newblock Census Tables.
\newblock Https://censusindia.gov.in/census.website/data/census-tables.

\bibitem[{Chanyaswad, Liu, and Mittal(2017)}]{chanyaswad2017ron}
Chanyaswad, T.; Liu, C.; and Mittal, P. 2017.
\newblock Ron-gauss: Enhancing utility in non-interactive private data release.
\newblock \emph{arXiv preprint arXiv:1709.00054}.

\bibitem[{Chow and Liu(1968)}]{chow1968approximating}
Chow, C.; and Liu, C. 1968.
\newblock Approximating discrete probability distributions with dependence trees.
\newblock \emph{IEEE transactions on Information Theory}, 14(3): 462--467.

\bibitem[{Cohen et~al.(2022)Cohen, Duchin, Matthews, and Suwal}]{cohen2022census}
Cohen, A.; Duchin, M.; Matthews, J.; and Suwal, B. 2022.
\newblock Census TopDown: The impacts of differential privacy on redistricting.
\newblock \emph{arXiv preprint arXiv:2203.05085}.

\bibitem[{Cohen and Nissim(2020)}]{cohen2020linear}
Cohen, A.; and Nissim, K. 2020.
\newblock Linear Program Reconstruction in Practice.
\newblock \emph{Journal of Privacy and Confidentiality}, 10(1).

\bibitem[{Ding et~al.(2021)Ding, Hardt, Miller, and Schmidt}]{ding2021retiring}
Ding, F.; Hardt, M.; Miller, J.; and Schmidt, L. 2021.
\newblock Retiring Adult: New Datasets for Fair Machine Learning.
\newblock \emph{Advances in Neural Information Processing Systems}, 34.

\bibitem[{Dwork et~al.(2006)Dwork, McSherry, Nissim, and Smith}]{dwork2006calibrating}
Dwork, C.; McSherry, F.; Nissim, K.; and Smith, A. 2006.
\newblock Calibrating noise to sensitivity in private data analysis.
\newblock In \emph{Theory of Cryptography: Third Theory of Cryptography Conference, TCC 2006, New York, NY, USA, March 4-7, 2006. Proceedings 3}, 265--284. Springer.

\bibitem[{Dwork and Roth(2014)}]{dwork2014algorithmic}
Dwork, C.; and Roth, A. 2014.
\newblock The algorithmic foundations of differential privacy.
\newblock \emph{Foundations and Trends{\textregistered} in Theoretical Computer Science}, 9(3--4): 211--407.

\bibitem[{Eurostat(2023)}]{eurostat_population_housing_censuses}
Eurostat. 2023.
\newblock Population and Housing Censuses Database.
\newblock Https://ec.europa.eu/eurostat/web/population-demography/population-housing-censuses/database.

\bibitem[{Fioretto, Van~Hentenryck, and Zhu(2021)}]{fioretto2021differential}
Fioretto, F.; Van~Hentenryck, P.; and Zhu, K. 2021.
\newblock Differential privacy of hierarchical census data: An optimization approach.
\newblock \emph{Artificial Intelligence}, 296: 103475.

\bibitem[{Ge et~al.(2020)Ge, Mohapatra, He, and Ilyas}]{ge2020kamino}
Ge, C.; Mohapatra, S.; He, X.; and Ilyas, I.~F. 2020.
\newblock Kamino: Constraint-aware differentially private data synthesis.
\newblock \emph{arXiv preprint arXiv:2012.15713}.

\bibitem[{gretelai(2023)}]{gretelrnn2023}
gretelai. 2023.
\newblock GretelRNN Implementation.
\newblock Https://github.com/gretelai/gretel-synthetics/tree/v0.15.10.

\bibitem[{Jordon, Yoon, and Van Der~Schaar(2018)}]{jordon2018pate}
Jordon, J.; Yoon, J.; and Van Der~Schaar, M. 2018.
\newblock PATE-GAN: Generating synthetic data with differential privacy guarantees.
\newblock In \emph{International conference on learning representations}.

\bibitem[{Liu and Wu(2022)}]{liu2022private}
Liu, T.; and Wu, Z.~S. 2022.
\newblock Private Synthetic Data with Hierarchical Structure.
\newblock \emph{arXiv preprint arXiv:2206.05942}.

\bibitem[{McKenna(2023)}]{ryan112358_private_pgm_mst}
McKenna, R. 2023.
\newblock MST Implementation.
\newblock Https://github.com/ryan112358/private-pgm/blob/master/mechanisms/mst.py.

\bibitem[{McKenna, Miklau, and Sheldon(2021)}]{mckenna2021winning}
McKenna, R.; Miklau, G.; and Sheldon, D. 2021.
\newblock Winning the NIST Contest: A scalable and general approach to differentially private synthetic data.
\newblock \emph{arXiv preprint arXiv:2108.04978}.

\bibitem[{McKenna, Sheldon, and Miklau(2019)}]{mckenna2019graphical}
McKenna, R.; Sheldon, D.; and Miklau, G. 2019.
\newblock Graphical-model based estimation and inference for differential privacy.
\newblock In \emph{International Conference on Machine Learning}, 4435--4444. PMLR.

\bibitem[{Narayanan and Shmatikov(2008)}]{narayanan2008robust}
Narayanan, A.; and Shmatikov, V. 2008.
\newblock Robust de-anonymization of large sparse datasets.
\newblock In \emph{2008 IEEE Symposium on Security and Privacy (sp 2008)}, 111--125. IEEE.

\bibitem[{Rosenblatt et~al.(2020)Rosenblatt, Liu, Pouyanfar, de~Leon, Desai, and Allen}]{rosenblatt2020differentially}
Rosenblatt, L.; Liu, X.; Pouyanfar, S.; de~Leon, E.; Desai, A.; and Allen, J. 2020.
\newblock Differentially private synthetic data: Applied evaluations and enhancements.
\newblock \emph{arXiv preprint arXiv:2011.05537}.

\bibitem[{{Singapore Department of Statistics}(2023)}]{singstat_tablebuilder}
{Singapore Department of Statistics}. 2023.
\newblock Singapore Statistics Table Builder.
\newblock Https://tablebuilder.singstat.gov.sg/.

\bibitem[{Tao et~al.(2021)Tao, McKenna, Hay, Machanavajjhala, and Miklau}]{tao2021benchmarking}
Tao, Y.; McKenna, R.; Hay, M.; Machanavajjhala, A.; and Miklau, G. 2021.
\newblock Benchmarking differentially private synthetic data generation algorithms.
\newblock \emph{arXiv preprint arXiv:2112.09238}.

\bibitem[{{United States Census Bureau}(2017)}]{kaggle_international_data}
{United States Census Bureau}. 2017.
\newblock International Datasets.
\newblock Https://www.kaggle.com/datasets/census/international-data.

\bibitem[{Vietri et~al.(2020)Vietri, Tian, Bun, Steinke, and Wu}]{vietri2020new}
Vietri, G.; Tian, G.; Bun, M.; Steinke, T.; and Wu, S. 2020.
\newblock New oracle-efficient algorithms for private synthetic data release.
\newblock In \emph{International Conference on Machine Learning}, 9765--9774. PMLR.

\bibitem[{Xie et~al.(2018)Xie, Lin, Wang, Wang, and Zhou}]{xie2018differentially}
Xie, L.; Lin, K.; Wang, S.; Wang, F.; and Zhou, J. 2018.
\newblock Differentially private generative adversarial network.
\newblock \emph{arXiv preprint arXiv:1802.06739}.

\bibitem[{Zhang et~al.(2017)Zhang, Cormode, Procopiuc, Srivastava, and Xiao}]{zhang2017privbayes}
Zhang, J.; Cormode, G.; Procopiuc, C.~M.; Srivastava, D.; and Xiao, X. 2017.
\newblock Privbayes: Private data release via bayesian networks.
\newblock \emph{ACM Transactions on Database Systems (TODS)}, 42(4): 1--41.

\bibitem[{Çağlar Tozluoğlu et~al.(2023)Çağlar Tozluoğlu, Dhamal, Yeh, Sprei, Liao, Marathe, Barrett, and Dubhashi}]{sweden2023pop}
Çağlar Tozluoğlu; Dhamal, S.; Yeh, S.; Sprei, F.; Liao, Y.; Marathe, M.; Barrett, C.~L.; and Dubhashi, D. 2023.
\newblock A synthetic population of Sweden: datasets of agents, households, and activity-travel patterns.
\newblock \emph{Data in Brief}, 48: 109209.

\end{thebibliography}

\end{document}